# Photoemission Study of the Electronic Structure of Valence Band Convergent SnSe

C. W. Wang[1,2,4,†], Y. Y. Y. Xia[2,3,4,†], Z. Tian[2,4,5,†], J. Jiang[2,6,7], B. H. Li[2], S. T. Cui[2], H. F. Yang[2], A. J. Liang[2], X. Y. Zhan[1,4], G. H. Hong[2,4,5], S. Liu[2,8], C. Chen[11], M. X. Wang[2], L. X. Yang[9,10], Z. Liu[1,2], Q. X. Mi[2], G. Li[2], J. M. Xue[2], Z. K. Liu[2*] and Y. L. Chen[2,9,11*]

[1]*Center for Excellence in Superconducting Electronics, State Key Laboratory of Functional Material for Informatics, Shanghai Institute of Microsystem and Information Technology, Chinese Academy of Sciences, Shanghai 200050, China*

[2]*School of Physical Science and Technology, ShanghaiTech University, CAS-Shanghai Science Research Center, Shanghai 200031, P. R. China*

[3]*Shanghai Institute of Applied Physics, Chinese Academy of Sciences, Shanghai 201800, China*

[4]*University of Chinese Academy of Sciences, Beijing 100049, China*

[5]*Shanghai Institute of Optics and Fine Mechanics, Chinese Academy of Sciences, Shanghai 201800, China*

[6]*Advanced Light Source, Lawrence Berkeley National Laboratory, Berkeley, CA 94720, USA*

[7]*Pohang Accelerator Laboratory, POSTECH, Pohang 790-784, Korea*

[8]*Shanghai Institute of Ceramics, Chinese Academy of Sciences, Shanghai 200050, China*

[9]*Collaborative Innovation Center of Quantum Matter, Beijing, P. R. China*

[10]*State Key Laboratory of Low Dimensional Quantum Physics, Department of Physics and Collaborative Innovation Center of Quantum Matter, Tsinghua University, Beijing 100084, P. R. China*

[11]*Department of Physics, University of Oxford, Oxford, OX1 3PU, UK*

†These authors contributed equally to this work.

*Corresponding authors: liuzhk@shanghaitech.edu.cn, yulin.chen@physics.ox.ac.uk

**IV-VI semiconductor SnSe has been known as the material with record high thermoelectric performance. The multiple close-to-degenerate (or "convergent") valence**

**bands in the electronic band structure has been one of the key factors contributing to the high power factor and thus figure-of-merit in the SnSe single crystal. Up to date, there has been only theoretical calculations but no experimental observation of this particular electronic band structure. In this paper, using Angle-Resolved Photoemission Spectroscopy, we performed a systematic investigation on the electronic structure of SnSe. We directly observe three predicted hole bands with small energy differences between their band tops and relatively small in-plane effective masses, in good agreement with the *ab-initio* calculations and critical for the enhancement of the Seebeck coefficient while keeping high electrical conductivity. Our results reveal the complete band structure of SnSe for the first time, and help to provide a deeper understanding of the electronic origin of the excellent thermoelectric performances in SnSe.**

Thermoelectric materials could directly convert heat (many times wasted) to electrical power and therefore are of critical importance in energy industry [1-7]. The conversion efficiency of thermoelectric materials is quantified by the dimensionless figure of merit, $ZT = S\sigma^2 T/\kappa$ ($S$: Seebeck coefficient, $\sigma$: electrical conductivity, $\kappa$: total thermal conductivity, including contributions from both electrons and phonons, $T$: temperature). Recently, single-crystalline SnSe, a binary IV-VI semiconductor compound containing non-toxic and earth-abundant elements, shows a record high *ZT* of ~2.6 at 923 K (along the b axis of the room-temperature orthorhombic unit cell) [8] and the device figure of merit ~1.34 from 300-773 K when hole-doped [9], much higher than that of typical high-performance thermoelectric materials [10-15]. These excellent thermoelectric performances can be attributed to both the relatively low thermal conductivity (~0.7 Wm$^{-1}$K$^{-1}$ at 300 K for the pristine samples) [8] as well as the very high Seebeck coefficient (~160 μVK$^{-1}$ at 300 K with carrier density of ~4×10$^{19}$ cm$^{-3}$) and power factor ($S\sigma^2$, ~40 μWcm$^{-1}$K$^{-2}$ at 300K) [9].

While the low thermal conductivity is attributed to the giant anharmonic and anisotropic bondings [8, 16, 17], the high Seebeck coefficient and power factor are deeply rooted in the electronic band structure of SnSe. It has been proposed that SnSe bears an electronic structure with relatively small effective mass (thus high mobility) [8, 18, 19] and multiple close-to-degenerate ("convergent") valence bands [9, 20, 21]. As the temperature increases, the carriers are thermally distributed over several convergent bands of similar energy, resulting in the enhanced Seebeck coefficient [22, 23]. Besides, the most outstanding electrical conductivity and power factor along the b axis among three axes of SnSe are thought to benefit from particular "pudding-mold-like" band [24-28]. However, although many theoretical calculations have predicted the unusual band structure as the electronic origin of excellent thermoelectric performance [9, 15, 18, 19, 24], experimental measurements and confirmation on band structure of SnSe remain unexplored.

In this paper, by carrying out Angle-Resolved Photoemission Spectroscopy (ARPES), we present a systematic study of the electronic band structure in single-crystalline SnSe. Our measurement has revealed the electronic band structure with band parameters similar to those of the *ab-initio* calculations. Especially, our ARPES data in the full Brillouin Zone (BZ) has identified "pudding-mold-like" band along the Γ-Y direction and a total of three valence bands with their band top energy differences <100 meV from each other, thus proving prominent band "convergence" in the electronic structure. Furthermore, we extracted the in-plane effective masses of all three valence bands, confirming that the effective masses of SnSe in the a-b plane are relatively small. Altogether, our measurement proves the convergent valence bands as well as the small effective mass all contribute to the excellent thermoelectric performance of single-crystalline SnSe, showing nice agreement with our *ab-initio* calculations.

High-quality SnSe single crystals were synthesized using the chemical vapor transport method. The orthorhombic crystal structure (with Pnma space group, #62) at room temperature is shown in Fig. 1(a). SnSe crystals host 2D sheets with strong, accordion-like corrugation. The sheets are well separated from each other and linked with weaker Sn-Se bondings [29-31], thus can be naturally cleaved along its [001] direction. The typical samples for our measurement have sizes around several millimeters (Fig. 1(b)(i)). The crystal structure is verified by the XRD diffraction pattern (Fig. 1(b)(ii)(iii)(iv)), from which we can extract the lattice constants as: a=4.15 Å, b=4.43 Å, c=11.50 Å, consistent with the previous reports [29-31]. From the topography image of the cleaved surface measured by Scanning Tunneling Microscope (STM) , we can confirm the natural cleavage plane is (001) termination without any observable reconstruction (see Fig. 1(c)). The X-Ray Photoelectron Spectrum (XPS) clearly shows the characteristic 4s, 4d peaks of Sn, and LMM, 3p, 3d peaks of Se (Fig. 1(d)), also suggesting the high quality of our samples.

We first describe the overall electronic structure of SnSe (001) surface. From the dispersions along the high symmetry $\bar{\Gamma} - \bar{X}$ and $\bar{\Gamma} - \bar{Y}$ directions we could observe serveral hole-like bands (Fig. 2(c), (d)). We could not observe the conduction band since SnSe is a semiconductor with the band gap of ~0.8-0.9 eV as suggested by the theoretical calculation and optical measuament and the Fermi level is sitting in the gap [8, 18, 19, 32]. Along the $\bar{\Gamma} - \bar{Y}$ direction, the α and β bands with the "pudding-mold-like" shape are identified with their band tops sitting around the same energy and forming the Valence Band Maximum (VBM). Around 0.07 eV below the VBM, another valence band (labelled as γ band) was observed between $\bar{\Gamma}$ and $\bar{X}$. Other hole-like bands are also observed with their band tops at least 0.5 eV below the VBM. The measured dispersions show overall agreement with the slab calculation results (Fig. 2(c), (d)), although the predicted ~10 meV energy difference between the band tops of α and β bands are not clearly observed probably due to the limited instrument resolution.

We confirm those band top positions by mapping the 3D electronic structure in the BZ. From the stacked plot of the electronic structure and constant energy contours (Fig. 2(a), (b)) we find only two features appear between $\bar{\Gamma}$ and $\bar{Y}$ ($k_x$=0, $k_y$=0.46 and 0.60 Å$^{-1}$) at the energy of VBM and an additional feature appears between $\bar{\Gamma}$ and $\bar{X}$ ($k_x$=0.47, $k_y$=0 Å$^{-1}$) around 0.07 eV below VBM. These features evolve into broad textures at higher binding energies, proving they are the band tops of α, β and γ bands, respectively. The mapping also proves that no other band tops are observed within ~0.1 eV below the VBM other than the α, β and γ bands in one full BZ. Our experimental spectra show a good agreement with the electronic structure of SnSe calculated from a tight-binding slab of 240 atomic layers (Fig. 2).

We investigate the evolution of the electronic structure along the $k_z$ direction by probing with different photon energies and comparing to the *ab-initio* calculation results (Fig. 3). From the dispersions along the high symmetry directions (Fig. 3(a), (b)) we could clearly observe several bands change rapidly with the photon energies. Especially, the "pudding-mold-like" α and β bands evolve from two separate bands with probing photon energy at 21 eV (corresponding to $k_z$=14π/c Å$^{-1}$) into one degenerate band with probing photon energies at 13 eV and 29 eV (corresponding to $k_z$=13 π/c and 15π/c Å$^{-1}$, respectively (Fig. 3(a)). Such evoltuion is apparent from the splitting and merging of the peaks in the Energy Distribution Curves (EDC) at $k_x=k_y=0$ (see the stacked line plot in Fig. 3(c)(i) and intensity plot in Fig. 3(c)(ii)). The apparent periodic band structure allows us to determine the $k_z$ values by comparison with the *ab-initio* calculation results. Besides, the top of the γ band becomes lowest when probed at 13 eV and 29 eV photon energies (corresponding to $k_z=13\pi/c$ and $15\pi/c$ Å$^{-1}$) and highest at 21 eV photon energy (corresponding to $k_z=14\pi/c$ Å$^{-1}$) (Fig. 3(b)). We note no other features besides α, β, γ are detected near the VBM and conclude these three bands dominate the electric and thermal transport behavior in SnSe when it is p-type doped. Our measurement of the electronic band structure at different $k_z$ values also shows good agreement with the *ab-initio* calculations (Fig. 3).

The observed band structure not only provides us information on the energy positions of the valence bands, but also the effective mass of each band. We extract the in-plane effective mass of the α, β, γ bands by fitting their dispersions around the band tops with parabolic curves and find the effective masses are $m^*_{kx}$=0.21 m$_0$, $m^*_{ky}$=0.13 m$_0$, $m^*_{kz}$=0.42 m$_0$ for the α band, $m^*_{kx}$=0.19 m$_0$, $m^*_{ky}$=0.12 m$_0$, $m^*_{kz}$=0.58 m$_0$ for the β band and $m^*_{kx}$=0.19 m$_0$, $m^*_{ky}$=0.13 m$_0$, $m^*_{kz}$=0.31 m$_0$ for the γ band (see Fig. 4 and supplemental material for detailed fitting processes, m$_0$ is the electron mass). The observed convergent valence band tops together with the small effective masses have profound effect in the thermal and electrical transport properties of SnSe. The small effective masses *m\** indicate the potential of obtaining high mobility holes, but on the other hand, they also give a small density of states (DOS), since $N_V \propto m^{*3/2}$, where $N_V$ is the DOS near the valence band top. A small $N_V$ is detrimental to achieving large Seebeck coefficient. In SnSe, this dilemma is naturally solved by the convergent valence band tops. With multiple peaks close in energy, although each single band effective mass could be small, together they contribute to a high effective mass and DOS. We believe that this is the electronic root of its high thermoelectric performance.

The observed electronic band structure helps us improve the theoretical model on the thermoelectric calculation of SnSe. The measured effective masses are smaller than the reported theoretical values within the Generalized Gradient Approximation (GGA) [9] and may lead to smaller Seebeck coefficients. The prominent agreement of the electronic structure between experiments and HSE06 calculations shown on Fig. 2-3 allows us to reexamine the validity of theoretical predictions on the thermoelectricity based on GGA calculations. To this end, we calculated Seebeck coefficients in both HSE06 and GGA approximations (the detailed comparison of our calculations and those in Ref [9] could be found in the supplemental material) at different carrier concentrations with the Boltztrap package [33]. The obtained Seebeck coefficients do not show much differences for heavily p-doped SnSe (carrier concentration $n = 1.6 \times 10^{20}$ cm$^{-3}$ and $3.0 \times 10^{20}$ cm$^{-3}$). In contrast, when the chemical potentials are close to the VBM ($n = 1.4 \times 10^{19}$ cm$^{-3}$ and $2.5 \times 10^{19}$ cm$^{-3}$), the Seebeck coefficients in HSE06 have smaller values comparing to those in GGA (Fig 5) due to the reduced effective mass. In addition, when the chemical potential is set above the VBM ($n = 5 \times 10^{17}$ cm$^{-3}$), the Seebeck coefficients display qualitatively difference in the two calculations: the curve starts to decrease at around 450 K and reverse sign at around 650 K in GGA calculation; while the one from HSE06 calculation monotonically increase until around 725 K, in better agreement with the previous experimental report [8]. This is due to the underestimated band gap in GGA calculation, where electrons are thermally excited across the gap at around 450 K, and therefore induces bipolar transport and sign change in the Seebeck coefficients with the increase of temperature. Since the validity of HSE06 method here is already verified by experiments, we conclude that the high Seebeck coefficient in Pnma-SnSe is obtainable with a light hole-doping in a wide range of temperature.

In conclusion, our combined ARPES measurement and *ab-initio* calculation reveals the complete electronic structure of SnSe. The nearly convergent α, β and γ bands, together with the small in-plane band effective masses, largely modify the Seebeck coefficients and thus the power factor, in particular at the VBM. Our work confirms the previous theoretical predictions and proposals that SnSe is a valence band convergent compound and hints the guideline searching for high *ZT* thermoelectric materials.

**Acknowledgements:**

We thank Dr. Zhe Sun for his assistance during the beamtime at NSRL; Dr. Makoto Hashimoto and Dr. Donghui Lu for their assistance during the beamtime at SSRL. This work is supported by grant from national key R&D program of China (2017YFA0305400) and Chinese Academy of Science-Shanghai Science Research Center (Grant No. CAS-SSRC-YH-2015-01).

Y.L.C. acknowledges the support from the Engineering and Physical Sciences Research Council Platform Grant (Grant No. EP/M020517/1). Z.K.L. acknowledges the support from the National Natural Science Foundation of China (11674229). Z. L. acknowledges the support from National Natural Science Foundation of China (11227902) and Science and Technology Commission of Shanghai Municipality (14520722100). All authors contributed to the scientific planning and discussions. The authors declare no competing financial interests.

**APPENDIX: MATERIALS AND METHODS**

**1. Sample synthesis:**

A fused silica tube (200 mm long, 22 mm inner diameter) was charged 2.50 g tin granules (99.999%) and 1.58 g selenium powder (99.999%), before being evacuated to $10^{-3}$ Pa and flame sealed. While keeping the mixture at one end, the whole tube was heated at 550 °C and then 880 °C, each for 12 h. Then the empty end of the tube was cooled over 1 h to 320 °C. After maintaining such a temperature gradient of 560 °C for 24 h, SnSe single crystals appeared on the wall of the cold end as grey flakes with metallic cluster.

**2. Angle-resolved photoemission spectroscopy**:

ARPES measurements were performed at beamline BL13U at National Synchrotron Radiation Laboratory (NSRL), China (photon energy hν = 13-29 eV) and beamline 5-4 at Stanford Synchrotron Radiation Lightsource (SSRL), USA (photon energy hν = 21 eV). The samples were cleaved *in situ*, and measured in ultrahigh vacuum with a base pressure of better than $3.5 \times 10^{-11}$ mbar, and data were recorded by a Scienta R4000 analyzer at a 20 K sample temperature. The energy/momentum resolution was 10 meV/0.2° (i.e., 0.006-0.009 Å$^{-1}$ for photoelectrons generated by 13-29 eV photons) for the NSRL setup, and 8 meV/0.2° (i.e., 0.008 Å$^{-1}$ for photoelectrons generated by 21 eV photons) for the SSRL setup.

**3. *Ab-initio* calculations:**

The *ab-initio* calculations were performed within the framework of density functional theory (DFT) as implemented in vienna ab initio simulation package (VASP) [34]. The lattice parameters and the atomic positions were optimized before the calculation of electronic structure. Heyd-Scuseria-Ernzerhof (HSE06) hybrid exchange functional [35] and a *k*-mesh of 12×12×5 were taken, and the kinetic energy cutoff was set to be 211.6 eV. As a comparison, DFT calculation within the generalized gradient approximation of Perdew-Burke-Ernzerhof [36] functional method was also performed.

The Boltztrap package was used in both GGA and HSE06 calculations to determine the corresponding Seebeck coefficients. From the converged charge and electronic structure, the group velocity $v_k$ of the state with energy $\epsilon_k$ was calculated which gives rise to the transport distribution $\Xi(\epsilon) = \sum_k \vec{v}_k \vec{v}_k \vec{\tau}_k$. The Seebeck coefficient was then obtained from the energy integral of $\Xi$, i.e. $S = \frac{ek_B}{\sigma} \int d\epsilon \left(-\frac{\partial f_0}{\partial \varepsilon}\right) \Xi(\epsilon) \left(\frac{\epsilon-\mu}{k_B T}\right)^2$, where $\sigma$ is the electrical conductivity $\sigma = e^2 \int d\epsilon \left(-\frac{\partial f_0}{\partial \varepsilon}\right) \Xi(\epsilon)$. The energy $\epsilon_k$ and group velocity $v_k$ were calculated in GGA and HSE06 as input to the Boltztrap. As in both calculations the same crystal structure was adopted, the change in effective mass is not due to the different bond lengths but from the different treatment of the electronic correlations, which is found to be crucial for achieving the agreement with ARPES on electronic structure.

**References:**

[1] G. J. Snyder and E. S. Toberer, Nature Materials **7**, 105 (2008).

[2] M. S. Dresselhaus, G. Chen, M. Y. Tang, R. G. Yang, H. Lee, D. Z. Wang, Z. F. Ren, J. P. Fleurial, and P. Gogna, Advanced Materials **19**, 1043 (2007).

[3] K. F. Hsu, S. Loo, F. Guo, W. Chen, J. S. Dyck, C. Uher, T. Hogan, E. K. Polychroniadis, and M. G. Kanatzidis, Science **303**, 818 (2004).

[4] L. E. Bell, Science **321**, 1457 (2008).

[5] F. J. Disalvo, Science **285**, 703 (1999).

[6] C. C. Li, F. X. Jiang, C. C. Liu, W. F. Wang, X. J. Li, T. Z. Wang, and J. K. Xu, Chemical Engineering Journal **320**, 201 (2017).

[7] H. Shi, W. Ming, D. S. Parker, M.-H. Du, and D. J. Singh, Physical Review B **95**, 195207 (2017).

[8] L. D. Zhao, S. H. Lo, Y. Zhang, H. Sun, G. Tan, C. Uher, C. Wolverton, V. P. Dravid, and M. G. Kanatzidis, Nature **508**, 373 (2014).

[9] L. D. Zhao *et al.*, Science **351**, 141 (2016).

[10] K. Biswas, J. He, I. D. Blum, C. I. Wu, T. P. Hogan, D. N. Seidman, V. P. Dravid, and M. G. Kanatzidis, Nature **489**, 414 (2012).

[11] P. F. Poudeu, J. D'Angelo, A. D. Downey, J. L. Short, T. P. Hogan, and M. G. Kanatzidis, Angewandte Chemie International Edition **45**, 3835 (2006).

[12] H. Z. Zhao *et al.*, Nano Energy **7**, 97 (2014).

[13] S. I. Kim *et al.*, Science **348**, 109 (2015).

[14] S. K. Plachkova, Physica Status Solidi (a) **83**, 349 (1984).

[15] H. J. Wu, L. D. Zhao, F. S. Zheng, D. Wu, Y. L. Pei, X. Tong, M. G. Kanatzidis, and J. Q. He, Nature Communications **5**, 4515 (2014).

[16] C. W. Li, J. Hong, A. F. May, D. Bansal, S. Chi, T. Hong, G. Ehlers, and O. Delaire, Nature Physics **11**, 1063 (2015).

[17] J. Carrete, N. Mingo, and S. Curtarolo, Applied Physics Letters **105**, 101907 (2014).

[18] G. S. Shi and E. Kioupakis, Journal of Applied Physics **117**, 065103 (2015).

[19] R. L. Guo, X. J. Wang, Y. D. Kuang, and B. L. Huang, Physical Review B **92**, 115202 (2015).

[20] L. D. Zhao, C. Chang, G. J. Tan, and M. G. Kanatzidis, Energy & Environmental Science **9**, 3044 (2016).

[21] K. L. Peng *et al.*, Energy & Environmental Science **9**, 454 (2016).

[22] A. Banik and K. Biswas, Journal of Solid State Chemistry **242**, 43 (2016).

[23] Y. Z. Pei, X. Y. Shi, A. LaLonde, H. Wang, L. D. Chen, and G. J. Snyder, Nature **473**, 66 (2011).

[24] K. Kutorasinski, B. Wiendlocha, S. Kaprzyk, and J. Tobola, Physical Review B **91**, 205201 (2015).

[25] K. Kuroki and R. Arita, Journal of the Physical Society of Japan **76**, 083707 (2007).

[26] Y. Nishikubo, S. Nakano, K. Kudo, and M. Nohara, Applied Physics Letters **100**, 252104 (2012).

[27] H. Usui, K. Suzuki, K. Kuroki, S. Nakano, K. Kudo, and M. Nohara, Physical Review B **88**, 075140 (2013).

[28] H. Usui and K. Kuroki, Journal of Applied Physics **121**, 165101 (2017).

[29] T. Chattopadhyay, J. Pannetier, and H. G. Vonschnering, Journal of Physics and Chemistry of Solids **47**, 879 (1986).

[30] W. J. Baumgardner, J. J. Choi, Y.-F. Lim, and T. Hanrath, Journal of the American Chemical Society 132, 9519 (2010).

[31] K. Adouby, C. Pérez-Vicen, J. C. Jumas, R. Fourcade, and A. A. Touré, Zeitschrift Fur Kristallographie 213, 343 (1998).

[32] M. Parenteau and C. Carlone, Physical Review B **41**, 5227 (1990).

[33] G. K. H. Madsen, D. J. Singh, Computer Physics Communications **175**, 67 (2006).

[34] G. Kresse and J. Furthmüller, Physical Review B **54**, 11169 (1996).

[35] J. Heyd, G. E. Scuseria, and M. Ernzerhof, Journal of Chemical Physics **118**, 8207 (2003).

[36] J. P. Perdew, K. Burke, and M. Ernzerhof, Physical Review Letters **77**, 3865 (1996).

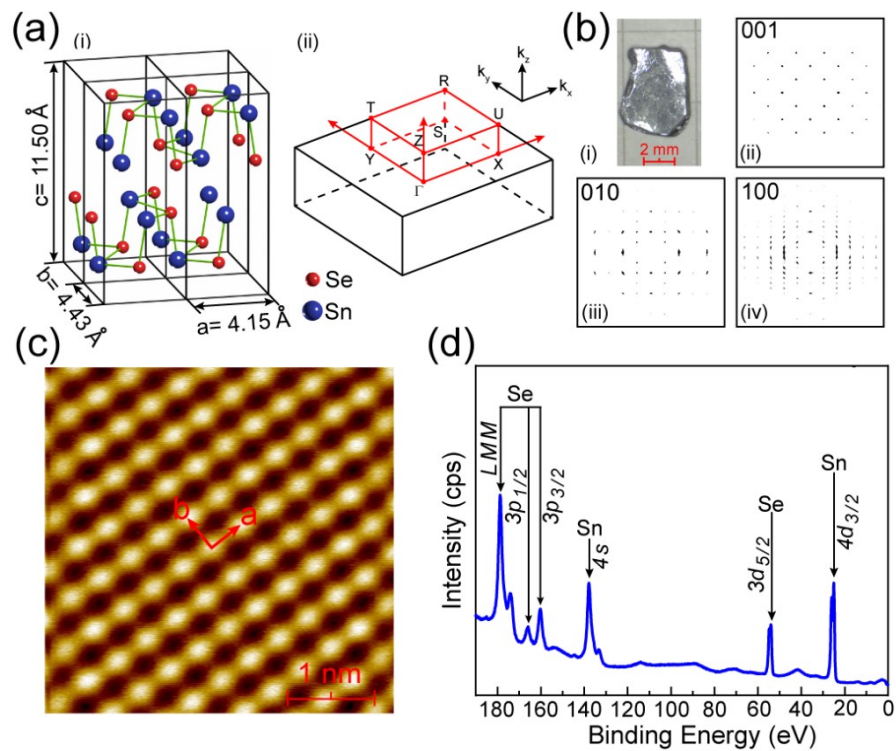

**FIG. 1. Crystal Structure and Characterization of SnSe.** (a) (i) Orthorhombic crystal structure of low-temperature SnSe, (ii) The corresponding BZ of SnSe, high symmetry points are labeled. (b) (i) Photograph of the high quality SnSe single crystal, (ii-iv) XRD pattern of the (001), (010), (100) surface of SnSe. (c) STM topography image of cleaved SnSe measured at ~77K. (d) The X-Ray Photoelectron Spectrum (XPS) of SnSe showing characteristic 4s, 4d core level peaks of Sn and LMM, 3p, 3d core level peaks of Se.

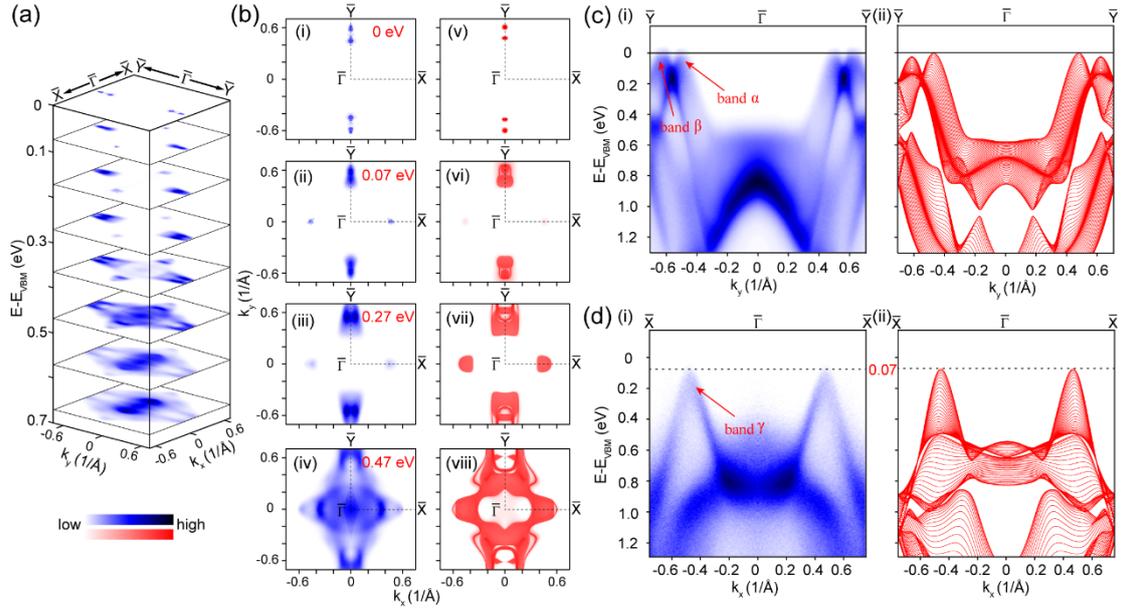

**FIG. 2. Basic Electronic Structure of SnSe.** (a) Stacking plots of constant energy contours in broad energy range showing the band structure evolution with high symmetry directions labeled. (b) (i-iv) Photoemission spectral intensity map showing constant energy contours at $E-E_{VBM}=$ 0, 0.07, 0.27 and 0.47 eV, respectively. (v-viii) Corresponding calculated constant energy contours at the same energies as (i-iv), respectively. (c) (i) High symmetry cut along the $\bar{Y} - \bar{\Gamma} - \bar{Y}$ with α, β band marked, (ii) Corresponding slab calculations along the $\bar{Y} - \bar{\Gamma} - \bar{Y}$ direction. (d) (i) High symmetry cut along the $\bar{X} - \bar{\Gamma} - \bar{X}$ direction with γ band marked, black dotted line indicates the top of the γ band. (ii) Corresponding slab calculations along the $\bar{X} - \bar{\Gamma} - \bar{X}$ direction. All data measured by 21 eV photons with linear horizontal polarization.

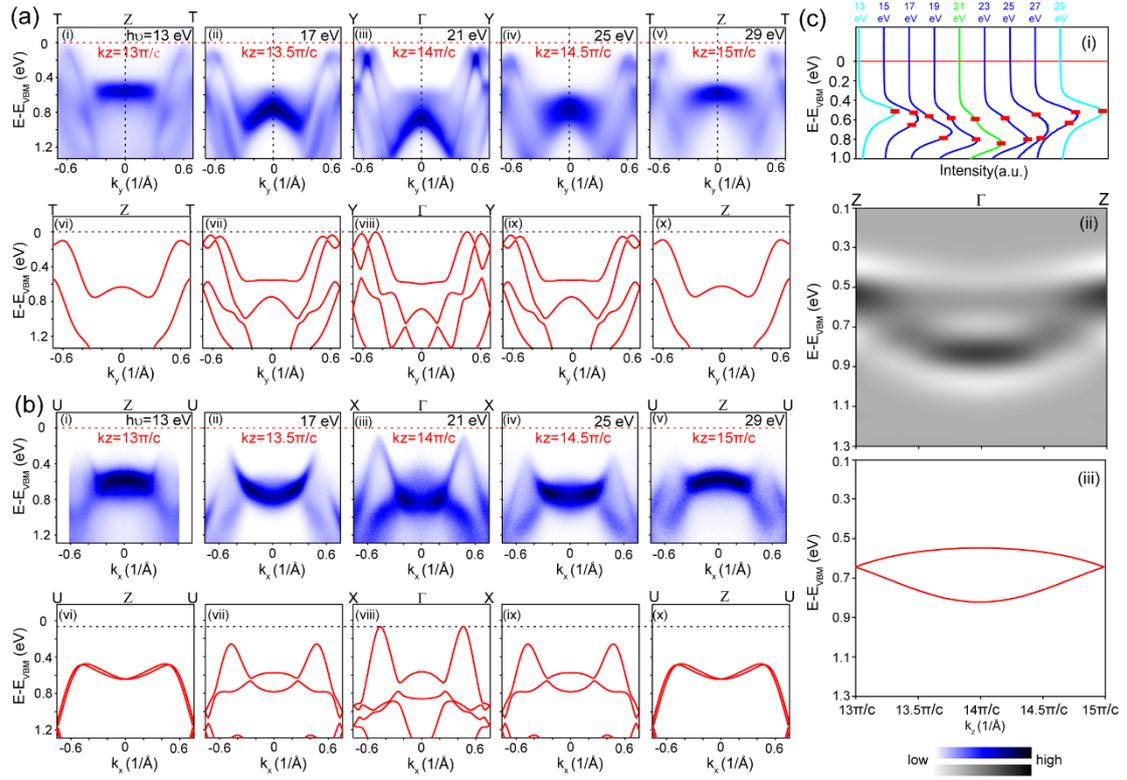

**FIG. 3. $k_z$ Evolution of the Electronic Bandstructure of SnSe.** (a) (i-v) High symmetry cuts along the $k_y$ direction ($k_x=0$) probed with different photon energies. Corresponding $k_z$ values of each dispersion are labeled. Perpendicular black dotted lines indicate the position ($k_x=k_y=0$) where the EDC is extracted and analyzed in (c)(i). (vi-x) Corresponding *ab-initio* calculated dispersions as in (i-v). (b) (i-v) High symmetry cuts along the $k_x$ direction ($k_y=0$) probed with different photon energies. Corresponding $k_z$ values of each dispersion are labeled. (vi-x) Corresponding *ab-initio* calculated dispersion as in (i-v). (c) (i) Stacked line plots of the Energy Distribution Curves (EDC) extracted at $k_x=k_y=0$ from different high symmetry cuts in (a)(i-v). Red marks indicate the identified peak positions. (ii) The second-derivative image of the intensity plot of (i) shows the dispersion along the Z-Γ-Z direction at $k_x=k_y=0$. (iii) Corresponding *ab-initio* calculated dispersion as in (ii). All data measured by photons with linear horizontal polarization.

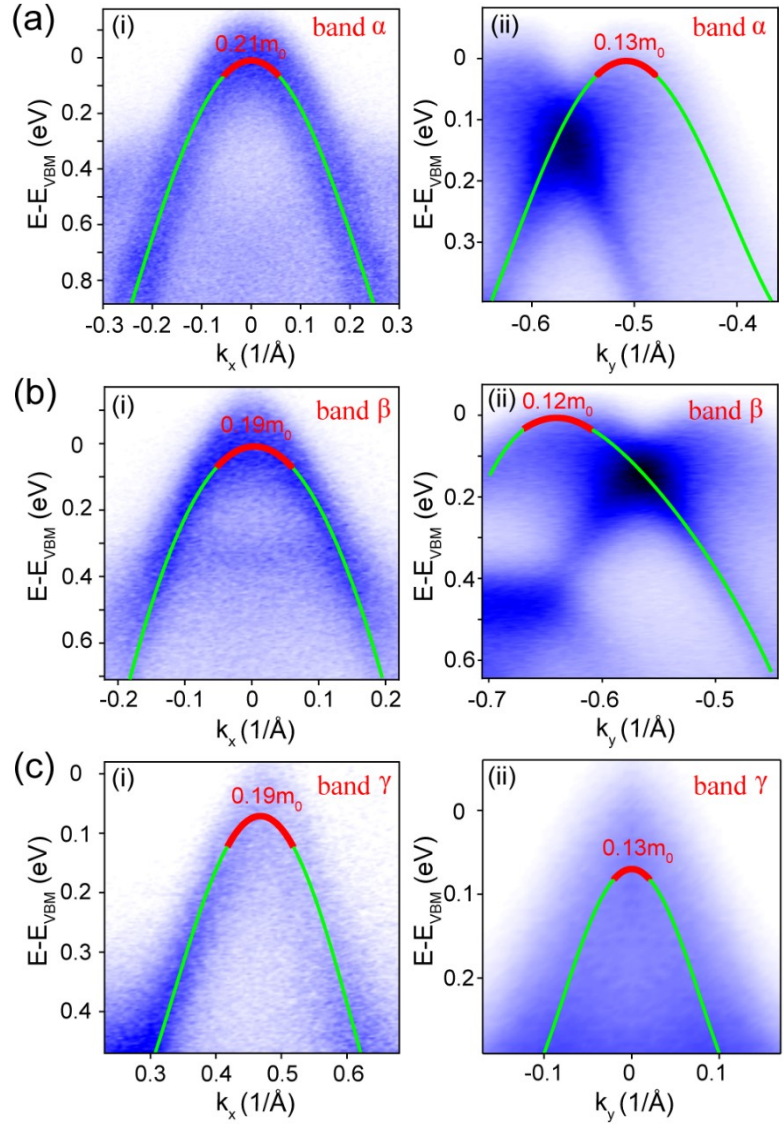

**FIG. 4. Extraction of the in-plane Effective Mass of Valence Bands in SnSe.** The in-plane effective masses of band α, β, γ along $k_x$ directions are extracted and labeled in (a)(i), (b)(i), (c)(i), respectively; those along the $k_y$ directions are extracted and labeled in (a)(ii), (b)(ii), (c)(ii), respectively. Green curves indicate a high order polynomial curve to fit the dispersion and red curves indicate the parabolic curves near the band top to extract the effective masses.

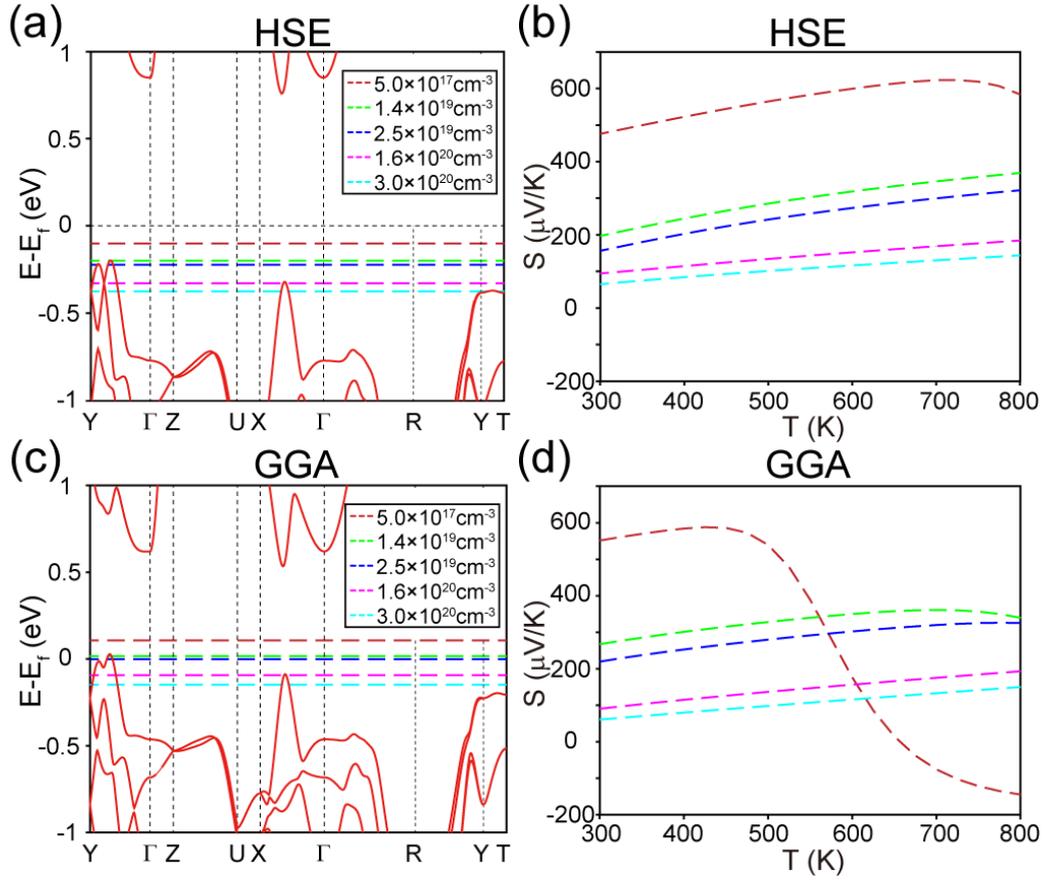

**FIG. 5. *Ab-initio* Calculation of the Seebeck Coefficient.** (a), (c) Band structure of Pnma-SnSe obtained from HSE06 (a) and GGA (c) calculations. The dashed lines from up to bottom correspond to the chemical potentials with the carrier concentrations of $5.0 \times 10^{17}$ cm$^{-3}$, $1.4 \times 10^{19}$ cm$^{-3}$, $2.5 \times 10^{19}$ cm$^{-3}$, $1.6 \times 10^{20}$ cm$^{-3}$, $3.0 \times 10^{20}$ cm$^{-3}$ at 300 K, respectively. (b), (d) The Seebeck coefficients from HSE06 and GGA calculations as a function of temperature, with respect to the chemical potentials with the same color in (a) and (c), respectively.